\documentclass[pre,twocolumn]{revtex4} 
\usepackage{amsmath,epsfig,amsfonts,amssymb}

\newcommand{\be}{\begin{equation}}
\newcommand{\ee}{\end{equation}}
\newcommand{\bp}{\begin{picture}}
\newcommand{\ep}{\end{picture}}
\newcommand{\ba}[1]{\begin{array}{#1}}
\newcommand{\ea}{\end{array}}
\newcommand{\bea}{\begin{eqnarray}}
\newcommand{\eea}{\end{eqnarray}}

\begin{document}

\title{Contact Angle of the Colloidal Liquid-Gas Interface and a Hard Wall}
\author{Paul P.\ F.\ Wessels}
\email{wessels@thphy.uni-duesseldorf.de}
\affiliation{Institut f\"{u}r Theoretische Physik II,
Heinrich-Heine-Universit\"{a}t 
D\"{u}sseldorf,
Universit\"{a}tsstra{\ss}e 1,
40225 D\"{u}sseldorf, Germany}

\author{Matthias Schmidt}
\altaffiliation[On leave from:]{
Institut f\"{u}r Theoretische Physik II,
Heinrich-Heine-Universit\"{a}t 
D\"{u}sseldorf,
Universit\"{a}tsstra{\ss}e 1,
40225 D\"{u}sseldorf, Germany}
\affiliation{Debye Institute, Utrecht University, 
Princetonplein 5, 3584 CC Utrecht, The Netherlands}

\author{Hartmut L\"{o}wen}
\affiliation{Institut f\"{u}r Theoretische Physik II,
Heinrich-Heine-Universit\"{a}t
D\"usseldorf,
Universit\"{a}tsstra{\ss}e 1,
40225 D\"{u}sseldorf, Germany}

\date{6 May 2004}

\begin{abstract}
We consider the Asakura-Oosawa-Vrij model of hard sphere colloids and
ideal polymer coils in contact with a planar hard wall at (colloidal)
liquid-gas coexistence. Using extensive numerical density functional
calculations, the liquid-gas, wall-liquid and wall-gas interfacial
free energies are calculated. The results are inserted into Young's
equation to obtain the contact angle between the liquid-gas interface
and the wall. As a function of polymer fugacity this angle exhibits
discontinuities of slope (``kinks'') upon crossing first-order surface
phase transitions located on the gas branch of the bulk binodal. Each
kink corresponds to a transition from $n-1$ to $n$ colloid layers
adsorbed at the wall, referred to as the $n$'th layering transition.
The corresponding adsorption spinodal points from $n-1$ to $n$
layers upon reducing the polymer fugacity along the bulk binodal were
found in a previous study (J.\ M.\ Brader et al.\ J.\ Phys.: Cond.\
Matt., 14: L1, 2002; Mol.\ Phys., 101: 3349, 2003).
Remarkably, we find desorption spinodal points from $n$ to $n-1$
layers to be absent upon increasing polymer fugacity at bulk
coexistence, and many branches (containing up to 7 colloid layers)
remain metastable. Results for the first layering binodal and both
spinodal branches off-bulk coexistence hint at a topology of the
surface phase diagram consistent with these findings. Both the order
of the transition to complete wetting and whether it is preceded by a
finite or an infinite number of layering transitions remain open
questions. We compare the locations of the first layering binodal line
and of the second layering binodal point at bulk coexistence with
recent computer simulation results by Dijkstra and van Roij (Phys.\
Rev.\ Lett., 89: 208303, 2002) and discuss our results for the contact
angle in the light of recent experiments.
\end{abstract}


\maketitle

\section{Introduction}
\label{sec:intro}

In suspensions of sterically-stabilized colloidal particles mixed with
nonadsorbing globular polymers, the latter induce an effective
attraction between the colloids due to the depletion
effect~\cite{poon02}.  Such mixtures can phase separate into two fluid
phases, one being a colloidal liquid that is rich in colloids and poor
in polymers and the other being a colloidal gas that is poor in
colloids and rich in polymers. Colloid-polymer mixtures serve as
excellent model systems to study many phenomena associated with
liquid-gas phase separation as time and length scales are much larger
than in atomic and molecular systems~\cite{poon02,lowen01}.  Recent
experiments have focused on the bulk phase behaviour~\cite{poon02}
(and Refs.\ therein), (colloidal) liquid-gas interface
tension~\cite{dehoog99a,dehoog99b,chen00,chen01,aarts03}, capillary
wave fluctuations observed in real space~\cite{aarts04b}, droplet
coalescence \cite{aarts04b} and further non-equilibrium
phenomena~\cite{poon02,aarts03,aarts04codef}. The behaviour at
nonadsorbing walls has been studied by measuring the contact angle
between the free liquid-gas interface and a substrate acting as a hard
wall. Complete wetting of the wall by the colloidal liquid has been
observed for a wide variety of statepoints ~\cite{aarts04codef}.
However, there are also reports of a transition from partial to
complete wetting~\cite{wijting03a,wijting03b}, hence this remains an
interesting topic.

Most of the essential physics of colloid-polymer mixtures is captured
by the Asakura-Oosawa-Vrij (AO) model of hard-sphere colloids and
ideal polymers~\cite{asakura54,asakura58,vrij76}, and which has become
a widely-used reference system.  Theoretical
approaches~\cite{gast83,lekkerkerker92} and computer
simulations~\cite{meijer91,meijer94,dijkstra99jpcm,bolhuis02,dijkstra02swet,vink03,vink04codef}
have given insight into its bulk phase behaviour, and some recent
work~\cite{dzubiella01a,bolhuis02} aims at including more realistic
colloid-polymer and polymer-polymer interactions. Studies based on a
one-component description of colloids interacting with an effective
depletion potential (obtained by integrating out the polymer degrees
of freedom and truncating at the pair-wise level) were devoted to
inhomogeneous situations, such as the free fluid-fluid interface
\cite{vrij97,brader00europhysLett} and adsorption at a hard wall
\cite{brader01inhom}.
Following the development of an accurate density functional theory
(DFT) specific for the binary AO model~\cite{schmidt00,schmidt02},
further research was stimulated in inhomogeneous situations such as
liquid-gas~\cite{brader02a,brader03molphys} and wall-fluid
interfaces~\cite{brader02a,brader03molphys,wessels04} and results were
compared to those from
simulations~\cite{dijkstra02swet,vink03,vink04codef,fortini04}.  In
particular, in Refs.~\cite{brader02a,brader03molphys} the AO model was
considered in contact with a planar hard wall. A sequence of
first-order layering transitions was found on the gas branch of the
(liquid-gas) binodal upon reducing the polymer fugacity. Further
reducing the polymer fugacity leads to a transition to complete
wetting of the wall by colloidal liquid.  This scenario was
corroborated by a simulation study~\cite{dijkstra02swet}. The relation
of the results from these different approaches will be reexamined in
the light of the findings of the present study in more detail below.
The adsorption properties at a wall are intimately related to the
wall-fluid interfacial free energies (or ``wall tensions''), for which
an analytical expression was obtained from a scaled-particle treatment
and which was found to compare well with results from full numerical
DFT calculations~\cite{wessels04}.  However, despite its experimental
accessibility~\cite{aarts03,aarts04codef,wijting03a,wijting03b}, the
contact angle of the free liquid-gas interface and a hard planar wall
has not been considered neither by theory nor simulations, in contrast
to wetting
behavior~\cite{brader02a,brader03molphys,dijkstra02swet,aarts04cahn}.
The aim of the present study is to obtain a quantitative understanding
of the contact angle and elucidate its relation to the surface phase
behavior on the basis of the AO model.

We obtain the (macroscopic) contact angle, $\theta$, from the
liquid-gas ($lg$), the wall-gas ($wg$) and the wall-liquid ($wl$)
interface tensions, $\gamma_{lg}$, $\gamma_{wg}$ and $\gamma_{wl}$,
respectively, via Young's equation~\cite{rw02},
\be 
\cos\theta =
\frac{\gamma_{wg}-\gamma_{wl}}{\gamma_{lg}}.
\label{eq:young}
\ee 
A similar depletion attraction that acts between two colloids acts
between one colloid and a hard wall \cite{brader01inhom}; the latter
therefore favours the colloidal liquid.  As a consequence, we expect
$\gamma_{wg}>\gamma_{wl}$ everywhere at coexistence, and
$\theta<\pi/2$.  For statepoints where
$\gamma_{wg}<\gamma_{wl}+\gamma_{lg}$, the contact angle $\theta >0$
and the surface is partially wet by the liquid.  However, as soon as
$\gamma_{wg}=\gamma_{wl}+\gamma_{lg}$ a macroscopic liquid layer will
intrude between the gas and the wall and the latter is completely wet
by the liquid.  The transition from partial to complete wetting
induced by changing an appropriate thermodynamic variable is referred
to as the wetting transition~\cite{sullivan86,dietrich88}.  A study of
$\theta$ can supply a link between theoretical predictions of surface
phase behavior and experiments and we display our central result in
Fig.~\ref{fig:contactanglecombo}.

The paper is organized as follows, in Sec.~II we define the model and
and discuss in Sec.~III the density functional theory used to
calculate the interface tensions.  In Sec.~IV we present results and
we conclude in Sec.V.

\section{Model}

We consider the Asakura-Oosawa-Vrij (AO) model of $N_c$ hard-sphere
colloids and $N_p$ ideal polymers in a volume $V$.  The colloids
(species $c$) and polymers (species$p$) have diameters $\sigma_i$,
bulk packing fractions $\eta_i=N_iV_i/V$, and particle volumes
$V_i=(\pi/6)\sigma_i^3$ for $i=c,p$, respectively.  The
colloid-colloid as well as the colloid-polymer interaction potentials
are those of hard spheres, so $u_{ij}(r)=\infty$ when
$r<(\sigma_i+\sigma_j)/2$ and $u_{ij}(r)=0$ otherwise, with
$ij=cc,cp$.  The polymers do not interact with each other, i.e.\
$u_{pp}(r)=0$ for all $r$.  Due to the non-additive ranges of these
interaction potentials, the polymers induce an effective attraction
between the colloids, which for sufficiently large size ratios
$\sigma_p/\sigma_c$ ($\gtrsim 0.35$) drives a thermodynamically stable
phase separation into a colloid-rich (liquid) and a colloid-poor (gas)
phase~\cite{gast83,lekkerkerker92}.  All bare interactions are of an
entropic nature, and therefore the temperature $T$ does not play a
role.  The only relevant model parameter is the size ratio
$q=\sigma_p/\sigma_c$.  We often use the so-called polymer-reservoir
representation where the mixture is in contact with a polymer
reservoir, which determines the polymer chemical potential.  In this
situation, the thermodynamic state parameters are $\eta_c$ and
$\eta_p^r$, the latter being the polymer packing fraction in the
reservoir which is proportional to the polymer fugacity as these
particles are ideal.

\section{Density Functional Theory}
We use the fundamental measure density functional for the AO model~\cite{schmidt00,schmidt02} to calculate colloid
and polymer density profiles from which the interface tensions can then be obtained.
In density functional theory (DFT), the grand-canonical free energy is expressed as a functional,
$\Omega[\rho_{c}(\mathbf{r}),\rho_{p}(\mathbf{r})]$, of the one-particle
distribution functions $\rho_i({\bf r})$ (with $i=c,p$),
given by~\cite{evans92}
\begin{multline}
\Omega[\rho_{c}(\mathbf{r}),\rho_{p}(\mathbf{r})]= 
F_{\rm exc}[\rho_{c}(\mathbf{r}),\rho_{p}(\mathbf{r})]\\ +
k_{\rm B} T \sum_{i= c,p}\int d\mathbf{r} \rho_i (\mathbf{r}) 
\left[\ln\left(\rho_i (\mathbf{r})\Delta_i\right)-1\right]  \\ +
\sum_{i= c,p}\int d\mathbf{r} \rho_i (\mathbf{r}) \left[ u_{{\rm ext},i}(\mathbf{r})-\mu_i\right],
\label{eq:grandfunctional}
\end{multline}
where $k_{\rm B}$ is Boltzmann's constant, $\Delta_i$ is the ``thermal
volume'' of species $i$, i.e.\ the third power of the de Broglie wave
length, and $u_{{\rm ext},i}(\mathbf{r})$ and $\mu_i$ are the external potential
and the chemical potential for species $i$, respectively.  
The excess Helmholtz free energy functional, $F_{\rm exc}=\int d{\bf r}\Phi({\bf
r})$, with $\Phi$ the excess free energy density, is given in
Refs.~\cite{schmidt00,schmidt02} and is not reproduced here%
~\footnote{We do not include the tensorial weight function in the
present calculations.}.

In thermodynamic equilibrium, the functional is stationary, $\delta\Omega/\delta\rho_i({\bf r})=0$ (with $i=c,p$), 
and the resulting equations yield the stable distributions, i.e.\
\be
\rho_{i}(\mathbf{r})=z_i \exp\left[-\beta u_{{\rm ext},i}(\mathbf{r}) -\beta \frac{\delta F_{\rm exc}[\{\rho_j 
(\mathbf{r})\}]}{\delta\rho_i (\mathbf{r})} 
\right],\label{eq:eulerlagrange}
\ee
with $z_i=\Delta_i^{-1}\exp[\beta \mu_i]$ the fugacity of component
$i$ and $\beta=1/k_{\rm B}T$.  
The equilibrium distribution functions are normalized, $\int d\mathbf{r}\rho_i (\mathbf{r})=N_i$.
We note that the polymer fugacity is
proportional to the polymer packing fraction in the polymer reservoir,
$\eta_p^r=z_pV_p$, and we usually refer to $\eta_p^r$ as the
polymer fugacity.  We have solved these equations numerically for
$\rho_i(z)$ in one spatial dimension $z$ for the free liquid-gas
interface (where $u_{{\rm ext},i}(z)=0$ everywhere) as well as both for the liquid and
the gas at bulk coexistence in the presence of the external hard-wall
potential, i.e.\ for $u_{{\rm ext},i}(z)=\infty$ for $z<\sigma_i/2$
and $u_{{\rm ext},i}(z)=0$ otherwise for both species $i=c,p$, where
$z$ is the space coordinate perpendicular to the wall.  The numerical
routine we have used is a Picard iteration procedure with a Broyles
mixing scheme~\cite{broyles60}.  The ``mixing parameter'' is continuously
adapted to obtain optimal convergence.  Additionally, it is important
to realize that in situations with several metastable minima, as we
find to occur for the coexisting gas in contact with the hard wall,
the initial guess for the profiles in the iteration procedure
determines to which minimum the routine converges.

Once the density profiles are known, the interface tension is given
by $\gamma=(\Omega_{\rm inh}+PV)/A$, where $\Omega_{\rm
inh}=\Omega[\rho_c(\mathbf{r}),\rho_p(\mathbf{r})]$ (i.e.\ the
functional, Eq.\ \ref{eq:grandfunctional}, evaluated at the solutions of
Eq.\ \ref{eq:eulerlagrange}) is the grand-canonical free energy of the
inhomogeneous system, $P$ is the bulk pressure, and $A$ is the lateral
(perpendicular to the $z$-direction) system area.  In terms of density
profiles this quantity can be written as
\be
\gamma=\int dz\left[ \omega (z)+P\right],
\label{eq:gamma}
\ee
where
\begin{multline}
\omega (z)=k_{\rm B}T\sum_{i=c,p}\rho_i(z)\left[\ln\left(\rho_i(z)\Delta_i\right)-1\right] 
\\ -\sum_{i=c,p}\mu_i\rho_i(z)+ k_{\rm B}T  \Phi(z)
\label{eq:gpdensity}
\end{multline}
can be viewed as a local grand potential density (evaluated with the
minimized density profiles).  In case of the liquid-gas interface, the
integral in Eq.~\ref{eq:gamma} is over all space, i.e.\ $z$ runs from
$-\infty$ (bulk gas) to $\infty$ (bulk liquid).  In case of the fluid
in contact with a hard wall, the integral runs from $z=0$ (at the
actual location of the hard wall) to $\infty$ (bulk).  In the
numerical routine, we compute the interface tensions for each
statepoint for different system sizes and refined termination criteria
for the iteration.  This gives an estimate of the numerical error in
the result for the interface tensions, which is important as the
resulting contact angle can be very sensitive to these errors.  Often
the dividing surface~\cite{rw02} is chosen at $z=\sigma_c/2$; then
subtraction of $\sum_{i=c,p}P\sigma_i$ from the present definition of
the wall-fluid tension, Eq.~\ref{eq:gamma}, is required. For our
present goal this is irrelevant as this term does not affect $\theta$,
as it drops out of the numerator in Eq.\ \ref{eq:young}.

\section{Results}

\subsection{Layered states at the hard wall}

\begin{figure}[t]
\epsfig{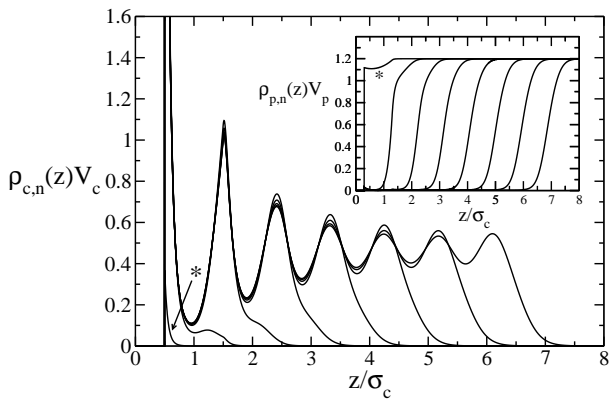}
\caption{\small Colloid density profiles, $\rho_{c,n}(z)V_c$, at a hard
wall as a function of the scaled distance, $z/\sigma_c$, from the wall
at the gas branch of liquid-gas binodal for $q=0.6$ and
$\eta_p^r=1.2$. Shown are results for $n$-layer states with
$n=0,1,2,\ldots, 7$ (left to right).  All states are metastable except
for the globally stable $n=0$-layer state (see the small peak at
contact with the wall, marked with an asterisk). The inset shows the
corresponding polymer profiles (also from left to right, with asterisk marking
$n=0$-layer state), $\rho_{p,n}(z)V_p$, as a function of $z/\sigma_c$. The normalizations of
the density profiles are such that in bulk they reduce to the
packing fractions, $\lim_{z\rightarrow\infty}\rho_{i,n}(z)V_i =\eta_i$.  }
\label{fig:profileszp1p2}
\end{figure}

\begin{figure}[t]
\epsfig{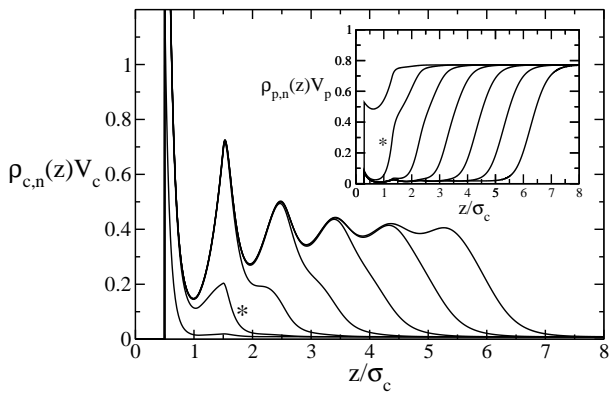}
\caption{\small Same as Fig.\ \ref{fig:profileszp1p2}, but for
$\eta_p^r=0.8$ and $n=0,1,2,\ldots, 6$ (left to right).  All profiles
are metastable except for the globally stable $n=1$-state (marked with
asterisks).}
\label{fig:profileszpp8}
\end{figure}

We have calculated colloid and polymer density profiles of the free
liquid-gas interface, as well as the wall-gas and the wall-liquid
interfacial profiles at bulk coexistence.  For a given statepoint the
liquid-gas and the wall-liquid profiles are both unique (disregarding
trivial translations of the liquid-gas interface); such results have
been presented elsewhere~\cite{brader02a,brader03molphys}.  However,
in case of the coexisting gas in contact with the hard wall, we have
found many metastable states, each corresponding to an integer number
of layers, $n$, of colloids adsorbed at the wall. In
Figs.~\ref{fig:profileszp1p2} and~\ref{fig:profileszpp8}, we have
plotted a number of such profiles, denoted by $\rho_{c,n}(z)$, for
size ratio $q=0.6$ and fugacities $\eta_p^r=1.2$ and $\eta_p^r=0.8$,
respectively, along with the corresponding polymer profiles,
$\rho_{p,n}(z)$, given in the respective insets (for reference, the
bulk phase diagram for $q=0.6$ is given in the inset of
Fig.~\ref{fig:surfacephaseqp6}).  For $\eta_p^r=1.2$ the 0-layer state
(marked with an asterisk in Fig.~\ref{fig:profileszp1p2}) exhibits
practically no excess colloid adsorption and is the globally stable
state.  The grand potential and hence the interface tension increases
with the number of layers, i.e.\
$\gamma_{wg,0}<\gamma_{wg,1}<\gamma_{wg,2}<\ldots<\gamma_{wg,7}$,
where $\gamma_{wg,n}$ is the wall-gas tension corresponding to $n$
colloid layers (as given in Eq.\ \ref{eq:gamma} and evaluated with
$\rho_{i,n}(z), i=c,p$). For $\eta_p^r=0.8$ the equilibrium profile is
given by $\rho_{i,1}(z)$ (marked with an asterisk in
Fig.~\ref{fig:profileszpp8}). Remarkably, in this case, the solution
$\rho_{i,0}(z)$ corresponds to a higher tension than all others, i.e.\
$\gamma_{wg,1}<\gamma_{wg,2}<\ldots<\gamma_{wg,6}<\gamma_{wg,0}$.  The
two state points considered ($\eta_p^r=0.8, 1.2$) are at polymer
fugacities larger than that at which the Fisher-Widom line
\cite{evans93,evans94} hits the bulk binodal (at $\eta_p^r\approx
0.533$ for $q=0.6$ \cite{schmidt02}), which implies that correlations
decay asymptotically in an oscillatory fashion in the liquid phase.
Apparently, this oscillatory nature also appears in the effective
interface potential between wall and the liquid-gas interface,
yielding many metastable minima~\cite{brader03molphys}, see 
Refs.\ \cite{henderson94}.

\subsection{Wall contact angle of the liquid-gas interface}

\begin{figure}[t]
\epsfig{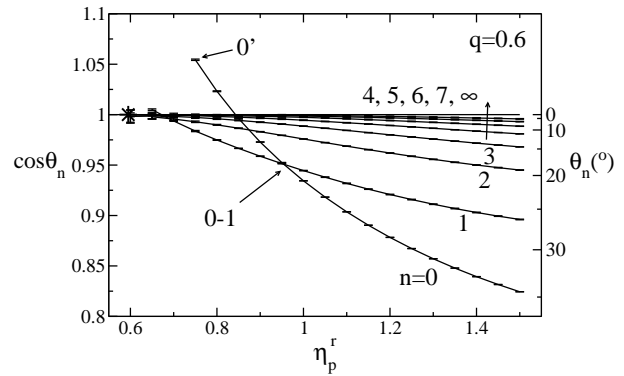}
\caption{\small Cosines of the contact angle, $\cos\theta_n$ as defined
via Young's equation as $(\gamma_{wg}-\gamma_{wl})/\gamma_{lg}$, as a
function of the polymer reservoir packing fraction, $\eta_p^r$, for
size ratio $q=0.6$. A scale of the bare angle (in degrees, $^{\rm o}$)
is given on the right vertical axis.  Shown are branches, $\theta_n$,
corresponding to $n$ colloid layers adsorbed at the wall-gas interface at
liquid-gas coexistence (with $n=0,1,2,\ldots,7$, increasing in the
direction of the arrow). The constant $\cos\theta_n =1$ ($\theta_n=0^{\rm
o}$) is marked $\infty$ and corresponds to two macroscopically
separated wall-liquid and liquid-gas interfaces.  For every value of
$\eta_p^r$, the branch corresponding to the lowest $\cos\theta_n$ (i.e.\
the largest contact angle $\theta_n$) yields the thermodynamically
stable state; all other branches are metastable. The crossing point
between $\theta_0$ and $\theta_1$ (marked 0-1) denotes the 
first layering transition
and the spinodal point of the 0-layer branch is marked 0'.
Also indicated is the location
of the wetting transition (star) according to Brader et
al.~\cite{brader02a,brader03molphys}.}
\label{fig:contactangleqp6}
\end{figure}

\begin{figure}[t]
\epsfig{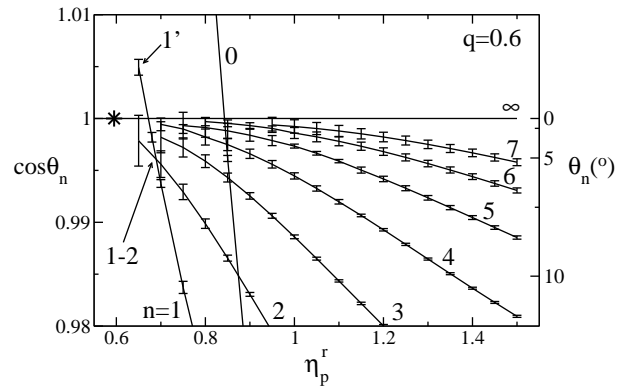}
\caption{\small Same as Fig.~\ref{fig:contactangleqp6}, but magnified
close to $\cos\theta_n=1$.  For the sake of clarity, some of the points
of the 3-, 4- and 5-layer branches close to the wetting transition are
omitted. Indicated is the location of the second
layering transition (marked 1-2) and the 1-layer spinodal point (1').}
\label{fig:contactangleqp6p}
\end{figure}

We have calculated the liquid-gas interface tension, $\gamma_{lg}$,
the wall-liquid tension, $\gamma_{wl}$, and the wall-gas tensions for
all $n$-layer states identified, $\gamma_{wg,n}$, for the full range
of polymer fugacities at bulk liquid-gas coexistence. The results have
been inserted into Young's equation, Eq.~\ref{eq:young}, yielding a
contact angle curve, $\theta_n$, for each $n$-layer state. The results
for $q=0.6$ are plotted in Fig.~\ref{fig:contactangleqp6} and a
magnification of the region close to $\cos\theta_n=1$ is displayed in
Fig.~\ref{fig:contactangleqp6p}. From the definition of the contact
angle, Eq.~\ref{eq:young}, and the fact that $\gamma_{lg}$ and
$\gamma_{wl}$ are unique for each statepoint, $\eta_p^r$, the state
$n$ with the lowest free energy also possesses the lowest value of
$\cos\theta_n$. Hence, for high $\eta_p^r$ the \emph{equilibrium}
contact angle is given by $\theta_0$ (corresponding to the 0-layer
state), see Fig.~\ref{fig:contactangleqp6}.  Decreasing $\eta_p^r$,
leads to an increase in $\cos\theta_0$, until it crosses the
$\cos\theta_1$ branch, and becomes metastable.  The crossing point,
where $\theta_0=\theta_1$, denotes the 0-1 layering transition and is
also referred to as the {\em first} layering transition.  (Consistent
with Refs.~\cite{brader02a,brader03molphys}, the surface phase
transition from $n-1$ to $n$ adsorbed colloid layers at the wall-gas
interface is referred to as the $n$'th layering transition.)  Upon
further decreasing $\eta_p^r$ (see Fig.~\ref{fig:contactangleqp6p}),
$\cos\theta_1$ is in turn crossed by $\cos\theta_2$ and the crossing
point, where $\theta_1=\theta_2$, is the 1-2 (second) layering
transition.  This scheme suggests that there could well be further
layering transitions upon reducing $\eta_p^r$, and this impression is
strengthened by the fact that the (metastable) states $\cos\theta_{n}$
(with $n$ from 3 to 7, see Fig.~\ref{fig:contactangleqp6p}), all seem
to converge to around the location of the wetting transition.
However, we have not been able to resolve the third
\cite{brader02a,brader03molphys} and (possible) higher layering
transitions and we note that any of these should be located in the
small region of $\cos\theta$ between 0.995 and 1 with $\eta_p^r$
between 0.6 and 0.65, see
Fig.~\ref{fig:contactangleqp6p}. Consequently, we can also not obtain
insight into the nature of the transition to complete wetting, i.e.\
whether this is second-order and occurs via an infinite sequence of
layering transitions or it is first order and is preceded by only a
finite number of layering transitions. For a more extensive discussion
of these two possible scenarios, we refer the reader to
Ref.~\cite{brader03molphys}.

Upon reducing $\eta_p^r$, metastable 0-layer states can be tracked
into a region where the contact angle takes unphysical values,
$\cos\theta_0 > 1$ (Fig.~\ref{fig:contactangleqp6}).  The same happens
for 1-layer states as can be seen in Fig.~\ref{fig:contactangleqp6p}.
Inserting equilibrium values of the interface tensions obtained from
DFT into Eq.~\ref{eq:young} ensures that $\cos\theta \leq 1$, but this
does not need to be the case when using interface tensions of
metastable states.  Reducing $\eta_p^r$ even further, the metastable
0-layer state eventually becomes unstable at an adsorption spinodal
point, see Fig.~\ref{fig:contactangleqp6}.  Beyond this point (for
even lower $\eta_p^r$), no 0-layer state can be stabilized and the
numerical iteration rather converges to the 1-layer state.  Similar
adsorption spinodal points were found for higher $n$ and we have
located those with a (moderate) resolution of 0.05 in $\eta_p^r$.
>From the present data the adsorption spinodal fugacities for $n$-layer
states, $\eta^r_{p,n'}$, are $\eta^r_{p,0'}=0.75$,
$\eta^r_{p,1'}=\eta^r_{p,2'}=0.65$ and
$\eta^r_{p,3'}=\eta^r_{p,4'}=\eta^r_{p,5'}=0.6$. For states with an
even thicker colloid film ($n=6$ and 7), the profiles no longer
converged properly at low $\eta_p^r$, and we can have not been able to
obtain precise values for $\eta^r_{p,6'}$ and $\eta^r_{p,7'}$.  The
evolution for surface states upon {\em increasing} $\eta_p^r$ is in
striking contrast. No spinodal points were found and each $n$-layer
state remains metastable up to $\eta_p^r=1.5$, a value close to the
liquid-gas-crystal triple point according to free-volume-theory for
the AO model~\cite{lekkerkerker92}. Moreover, for large values of
$\eta_p^r$ the numerical routine converges very rapidly, which hints
at deep (nevertheless metastable) free-energy minima for these layered
surface states.

As the layering transitions are thermodynamic surface phase
transitions, they manifest themselves as discontinuous jumps in the
Gibbs adsorption~\cite{brader02a,brader03molphys},
\be
\Gamma_i=\int_0^{\infty}{\rm d}z(\rho_i(z)-\rho_i(\infty)),
\label{eq:adsorptionFromProfile}
\ee
for both components $i=c,p$, and these can be obtained from Eq.~\ref{eq:gpdensity} and
\begin{equation}
\Gamma_c=-\left.\frac{\partial\gamma_{wf}}{\partial\mu_c}\right|_{\mu_p},\quad
\Gamma_p=-\left.\frac{\partial\gamma_{wf}}{\partial\mu_p}\right|_{\mu_c},
\label{eq:adsorptionDefintion}
\end{equation}
where $\gamma_{wf}$ is the wall tension of the fluid. 
Moving along the gas branch of the liquid-gas bulk binodal 
ties together changes in both chemical potentials:
\begin{equation}
 \left.\frac{d\gamma_{wg}}{d\mu_p}\right|_{\rm coex} =
 \left.\frac{\partial\gamma_{wg}}{\partial\mu_p}\right|_{\mu_c}
 +\left.\frac{\partial\gamma_{wg}}{\partial\mu_c}\right|_{\mu_p}
 \left.\frac{d\mu_c}{d\mu_p}\right|_{\rm coex},
\end{equation}
where the slope of the bulk binodal fulfills a Clapeyron-type
equation,
\begin{equation}
 \left.\frac{d\mu_c}{d\mu_p}\right|_{\rm coex} = 
 -\frac{\Delta\rho_p}{\Delta\rho_c},
\end{equation}
which can be deduced from the Gibbs-Duhem equation in a straightforward fashion~\cite{schmidt04aog}.
Here, $\Delta\rho_i=\rho_i^l-\rho_i^g$ is the difference in density
of species $i=c,p$ in the liquid and in the gas phase (note that
$\Delta\rho_p<0$). Hence, we obtain
\begin{equation}
 \left.\frac{d\gamma_{wg}}{d\mu_p}\right|_{\rm coex}  = 
 -\Gamma_{p} + \Gamma_{c} \frac{\Delta\rho_p}{\Delta\rho_c},
 \label{eq:slope}
\end{equation}
where the adsorptions $\Gamma_c$ and $\Gamma_p$ refer to those of the
gas at bulk coexistence.  Hence, as crossing a layering transition
(again at bulk coexistence) is necessarily accompanied by a jump in
the adsorptions $\Gamma_{i}$, consequently via Eq.~\ref{eq:slope} this
leads to a discontinuity of slope of the wall-gas interface tension.
>From Young's equation~\ref{eq:young}, it follows that this also leads
to a jump in the slope of the contact angle,
$d\cos\theta/d\eta_p^r$. This is consistent with our findings above of
crossing of different branches, $\theta_{n-1}$ and $\theta_n$
\footnote{The reported jump in $\gamma_{wg}$ as a function of
difference in colloid packing fractions in the liquid and gas phases
at the first layering transition \cite{wessels04} arises from crossing
the adsorption spinodal rather than the binodal. The equilibrium curve
for $\gamma_{wg}$ is continuous.}.

Note that all quantities in Eq.~\ref{eq:slope} can be independently
obtained from our DFT results, namely the adsorption $\Gamma_i$ from
the integral over the respective density profile, 
Eq.~\ref{eq:adsorptionFromProfile}, the differences $\Delta\rho_i$ from
the bulk phase diagram, and the left hand side of Eq.~\ref{eq:slope}
from a numerical derivative of the results for $\gamma_{wg}$, as
obtained through Eq.~\ref{eq:gamma}.  As a check for internal
consistency of our calculations we have chosen the statepoint with
$\eta_p^r=1.1$ for $q=1$ at bulk coexistence, which is very close to
the binodal of first layering transition (which is at $\eta_p^r\approx
1.104$, as discussed below). We find Eq.~\ref{eq:slope} to be
fulfilled to three significant digits; for the 0-layer (1-layer) state
either sides evaluate to $1.264/\sigma^2_c$ ($1.352/\sigma^2_c$). 
Together these two estimates yield a jump in $d\cos\theta/d\eta_p^r$ of 0.247,
which is consistent with our data for the contact angle for $q=1$ (which
gives 0.238) and which is discussed below.


\begin{figure}[t]
\epsfig{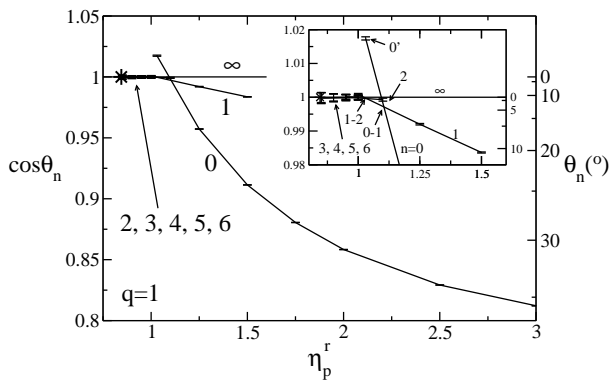}
\caption{\small Same as Fig.~\ref{fig:contactangleqp6}, but for
$q=1$. The inset displays a magnification of the area close to the
wetting transition. The points marked 0-1 and 1-2 are the
first layering and second layering transitions, respectively,
and spinodal point of the 0-layer branch is denoted with 0'.  }
\label{fig:contactangleq1}
\end{figure}

We will now consider the case of larger polymer-colloid size ratios,
$q=1$, see Fig.~\ref{fig:contactangleq1} in more detail.  The 0-layer
branch is calculated for a number of fugacities from high values,
$\eta_p^r=3$, to the spinodal point close to $\eta_p^r=1$.  The
higher-$n$ branches are only calculated between 0.95 and 1.5 ($n=1$)
and between the wetting transition~\cite{brader03molphys} and the
crossing point of the 0-layer and the 1-layer branch ($n=2-6$, which
practically fall on top of each other).  It is found that for this
size ratio the 0-layer branch is the stable everywhere except in a
small regime of $\eta_p^r$ between 0.85 and 1.1 (where $\cos\theta_n$
is close to 1) and where layering transitions are located.  We have
been able to identify the locations of the first and second layering
transitions, higher transitions are eroded by the numerical noise.

\begin{figure}[t]
\epsfig{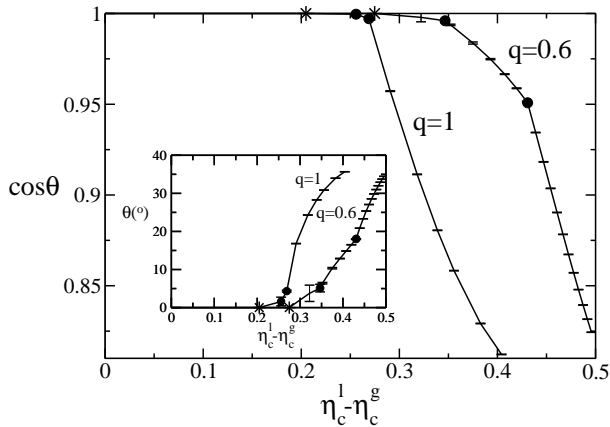}
\caption{\small The equilibrium contact angle, $\cos\theta$ (main
figure) and $\theta$ (in degrees, displayed in the inset), as a
function of the difference in colloid packing fractions,
$\eta_c^l-\eta_c^g$, of the coexisting bulk liquid and gas phases for
two size ratios, $q=0.6$ and $q=1$. 
The locations of the equilibrium layering transitions are marked by filled circles.
The stars are the transitions from partial to complete wetting according to Brader
et al.~\cite{brader02a,brader03molphys}.}
\label{fig:contactanglecombo}
\end{figure}

Fig.~\ref{fig:contactanglecombo} shows the resulting equilibrium
values of $\cos\theta$ (see the inset for the bare angle $\theta$) as
a function of the difference in colloid packing fractions of the
coexisting liquid and gas phases, $\eta_c^l-\eta_c^g$, for both size
ratios considered, $q=0.6$ and $q=1$. This representation enables one
to make direct contact with experiments, where density differences of
coexisting phases are rather directly
accessible~\cite{aarts03,aarts04codef,wijting03a,wijting03b}. For both
size ratios the contact angle is of the same order of magnitude, i.e.\ 
$\cos\theta=0.8-1$, but for
$q=1$ the wetting transition lies closer (in this representation) to
the bulk critical point.  Only the first layering transition for
$q=0.6$ occurs at a considerably large contact angle of
$\cos\theta_{0-1}\approx 0.95$ or $\theta_{0-1}\approx 18^{\rm o}$. A
remarkable fact is that for both size ratios, there is a substantial
region in the partial wetting regime close to the wetting transition
where the contact angle remains very small, i.e.\ in the range
$(\eta_c^l-\eta_c^g)\approx 0.28-0.35$ for $q=0.6$ and
$(\eta_c^l-\eta_c^g)\approx 0.2-0.27$ for $q=1$.

\subsection{Surface phase behaviour}

\begin{figure}[t]
\epsfig{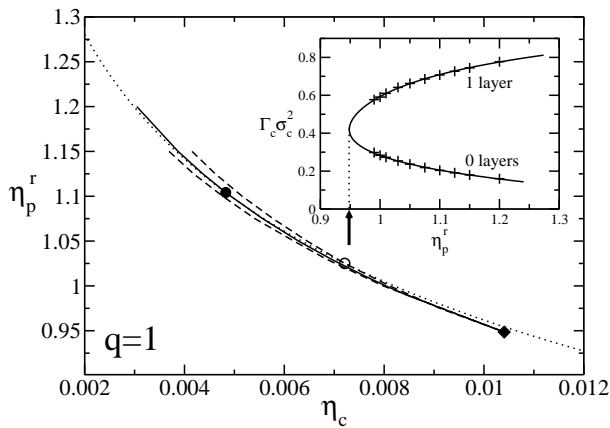}
\caption{\small The full surface phase diagram of the first layering
transition (from the 0-layer to the 1-layer state) for $q=1$ as a
function of (bulk) colloid packing fraction, $\eta_c$, and polymer
reservoir packing fraction, $\eta_p^r$. Shown is the layering binodal
(full line), the 1-layer desorption spinodal (dashed line, left),
0-layer adsorption spinodal (dashed line, right), and the gas branch
of the bulk liquid-gas binodal (dotted line).  The point where the
layering binodal crosses the bulk liquid-gas binodal is the first
layering transition at bulk coexistence (filled circle).  The point
where the 0-layer adsorption spinodal line hits the bulk binodal is
the 0-layer spinodal point at bulk coexistence (open circle),
according to Brader et al.~\cite{brader03molphys}. Both spinodals and
the binodal terminate at the layering critical point (filled diamond)
located off-bulk coexistence in the one-phase gas region.  Note that
the 1-layer desorption spinodal runs entirely in the one-phase gas
region and hence does not cross the bulk binodal. 
The inset shows the colloid adsorption, $\Gamma_c\sigma_c^2$, of 
the coexisting 0-layer and 1-layer states (along the layering binodal) 
as a function of polymer reservoir packing fraction, $\eta_p^r$;
the dividing surface is chosen at $z=\sigma_c/2$ (lower bound of
the integral in Eq.\ \ref{eq:adsorptionFromProfile}).
The symbols are obtained from DFT, the full line is the fit (see text) 
and the arrow at the horizontal axis denotes the
estimated value of $\eta_p^r$ at the layering critical point.}
\label{fig:1stlayeringq1}
\end{figure}

Layering transitions of the AO model at a hard wall were reported
earlier by Brader et al.~\cite{brader02a,brader03molphys} as obtained
within the same DFT approximation \cite{schmidt00,schmidt02} and using
numerical routines similar to those employed in the present
work. However, in contrast to the present study, Brader et al.\
determined (implicitly) spinodal points along two paths: First, by
reducing $\eta_p^r$ at bulk (liquid-gas) coexistence and following the
evolution of the 0-layer state, observing jumps to 1-layer and
subsequently higher-layer states, and second, by keeping $\eta_p^r$
fixed and approaching the gas branch of the binodal by increasing
$\eta_c$, starting from very small values.  Results along reversed
paths for the second case, i.e.\ decreasing $\mu_c$ at constant $\eta_p^r$, were taken to
ascertain that hysteresis effects are small, and it was concluded that
the spinodal points give reasonable indications of the locations of
the equilibrium layering
transitions~\cite{brader03molphys,brader04priv}.  We will discuss in
the following the relation of these findings to those of the present
work.
For this purpose, we have chosen a reference case, i.e.\ the first
layering transition for $q=1$, and mapped out its surface phase behaviour completely,
i.e.\ including binodal and spinodal lines off-bulk coexistence.
The result is plotted in Fig.~\ref{fig:1stlayeringq1}.

The layering transition binodal is obtained by determining, for each
value of $\eta_p^r$, the value of $\eta_c$ at which the 0-layer and
1-layer states have equal interface tensions and are thus in
thermodynamic coexistence. The resulting binodal line extends below
the first layering transition at bulk coexistence, at
$\eta^r_{p,0-1}\approx 1.1$, to lower values of $\eta_p^r$ into the
bulk gas phase region.  As this is a transition between two (surface)
phases of the same symmetry, a van-der-Waals loop in the free energy
and a critical point are mandatory. In order to find the location of
the critical point, we fit our data for the colloid adsorption
$\Gamma_c(\eta_p^r)$ of the coexisting 0-layer and 1-layer states with
a fourth-order polynomial, $\eta_p^r=a_0 + a_2(\Gamma_c-a_1)^2+
a_3(\Gamma_c-a_1)^3+ a_4(\Gamma_c-a_1)^4$, where the $a_i$ are free
fit parameters, see Fig.~\ref{fig:1stlayeringq1}.  The value of $a_0$
is an estimate of the critical value of $\eta_p^r$, and the functional
form is chosen to yield the mean-field critical exponent of 1/2, i.e.\
$(\Gamma_c-a_1)\sim(\eta_p^r-a_0)^{1/2}$, as the DFT is a mean-field
theory in the sense that it does not capture fluctuation effects.  The
accuracy in colloid packing fraction $\eta_c$ of points on the
layering binodal is high; typically smaller than 0.1\% of $\eta_c$.
However, the critical point is an extrapolation and is therefore much
more sensitive to errors, i.e. these may be as large as 3\% in
$\eta_p^r$ and 1\% in $\eta_c$.  We have also located the 0-layer
spinodal by taking paths at constant $\eta_p^r$ and increasing values
of $\eta_c$ monitoring the stability of the solution under the
iteration procedure. The values of $\eta_c$ where the 0-layer state
becomes unstable and converges to the 1-layer state defines the
``adsorption spinodal'', located at {\em larger} values at $\eta_c$
compared to the layering binodal. (The precise location of spinodal
points is subject to the numerical resolution of the step size in
$\eta_c$, i.e.\ typically 1\% in $\eta_c$.)  Similarly, we have
investigated the stability of the 1-layer states upon reducing
$\eta_c$ at constant $\eta_p^r$. This defines the ``desorption
spinodal'', where the 1-layer solutions converge to the 0-layer states
and which is located at {\em smaller} values of $\eta_c$ as compared
to the layering binodal. Upon decreasing $\eta_p^r$, both spinodals
and the layering binodal end at the surface critical point. Indeed
even for values of $\eta_p^r$ quite above the layering critical point
the adsorption and desorption spinodals are very close and we can
confirm the finding of Brader et al.\ that hysteresis effects at
constant $\eta_p^r$ are small.

Next, we discuss the various aspects of this surface phase transition
in relation to bulk liquid-gas coexistence.  We first note that no
intrinsic difference is observed between the layering phase transition
in the stable gas region and that in the two-phase region where the
gas is metastable.  In the following, we consider the three different
surface phase transition lines, i.e.\ the adsorption spinodal,
layering binodal and desorption spinodal, and their relation to the
bulk binodal. First, the crossing point of the adsorption spinodal and
the bulk liquid-gas binodal denotes the spinodal point terminating the
metastability region of the 0-layer state upon reducing $\eta_p^r$ at
bulk coexistence.  This corresponds to the adsorption spinodal point
of the contact angle (see Fig.~\ref{fig:contactangleq1}), as located
previously by Brader et al.\ (as well as other adsorption spinodal
points at bulk coexistence for different size ratios,
$q=0.6,0.7,1$). Our numerical value of $\eta_p^r$ agrees well with
that of Ref.\ \cite{brader03molphys}. Second, the crossing point
between layering binodal and bulk binodal is the layering transition
at bulk coexistence (accompanied by a kink in the contact angle, as
outlined above). This statepoint can be seen as a triple point between
the bulk liquid and two different surface states of the bulk gas. The
location of this triple point is quite different from the adsorption
spinodal point at bulk coexistence.  Although hysteresis for a path at
constant $\eta_p^r$ is small, this is not the case for the path along
bulk coexistence, due to the fact that the gas branch of the bulk
binodal and the layering (spinodal and binodal) lines have very
similar slopes. Hence the location of any crossing point is very
sensitive to the precise location of the individual lines. Third, in
striking contrast to the previous two cases, the desorption spinodal
{\em does not} cross the bulk binodal, but remains in the one-phase
gas region for increasing values of $\eta_p^r$.  We have checked this
for one additional path at $\eta_p^r=2$, starting with a 1-layer
profile at bulk coexistence and decreasing $\eta_c$ and indeed found
the desorption spinodal point in the one-phase gas region.  This
behaviour is consistent with the behaviour of the contact angle, which
we discussed above for $q=0.6$ and $q=1$, and where adsorption
spinodal points were found upon decreasing $\eta_p^r$, but no
desorption spinodal points were found upon increasing $\eta_p^r$.
Although we have determined this scenario only extensively for the
first layering transition for $q=1$, we believe it to hold more
generally for higher layering transitions and other size ratios.

\begin{figure}[t]
\epsfig{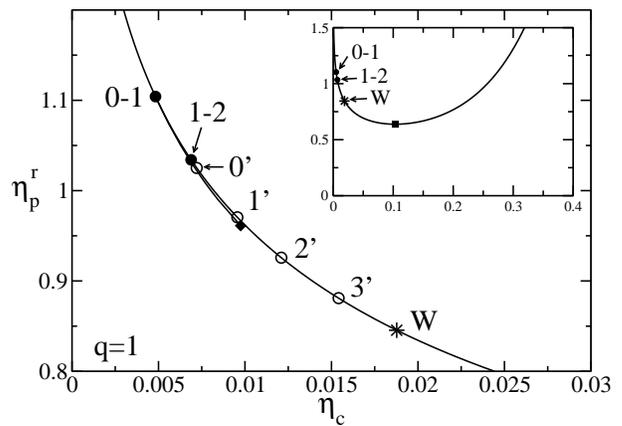}
\caption{\small Summary of the known features of the surface phase
diagram of the AO model at a hard wall for size ratio $q=1$ as a
function of colloid packing fraction, $\eta_c$, and polymer reservoir
packing fraction, $\eta_p^r$.  Shown are the gas branch of the bulk
binodal (full line), equilibrium first (filled circle, marked 0-1) and
second (filled circle, marked 1-2) layering transitions at bulk
coexistence, as obtained from the present work. The first layering
critical point (diamond) is connected via the layering binodal line
(full curve) to the first layering transition at bulk coexistence. The
spinodal lines are omitted for clarity.  The 0-, 1-, 2-, 3-layer
adsorption spinodal points at bulk coexistence (open circles, marked
$0'$, $1'$, $2'$ and $3'$ respectively) and the wetting transition (star, marked W)
are taken from Refs.~\cite{brader03molphys}). The inset shows part of the
data together with the bulk critical point (large filled square) on a
larger scale.}
\label{fig:surfacephaseq1}
\end{figure}

In order to summarize our present results and those of
Ref.~\cite{brader03molphys} for the layering transitions of the AO
model at a hard wall, we draw the surface phase diagram for $q=1$ in
Fig.~\ref{fig:surfacephaseq1}. The results for the $n$-layer
adsorption spinodal points for $n=0,1,2,3$ at bulk coexistence are
taken from Ref.~\cite{brader03molphys}. At bulk coexistence, the first
and the second layering binodal points are located at higher
fugacities compared to the corresponding 0-layer and 1-layer
adsorption spinodal points, respectively.  Higher layering
transitions, corresponding to 2- and 3-layer spinodal points of Brader
et al.\ do not emerge from our present data. On the other hand, Brader
et al.\ have not found layering lines extending into the bulk gas
phase for $q=1$ and which we have located for the first layering
transition.  We have not searched for a similar layering line in case
of the second layering transition. However, from the topology of the
first layering transition, which we established above, and taking into
account the considerable separation of the second layering binodal and
the 1-layer adsorption spinodal points at bulk coexistence, it seems
plausible that such a layering binodal line also exists for the second
layering transition.

Next, we compare our results to those from simulations by Dijkstra and
van Roij for $q=1$~\cite{dijkstra02swet}.  They find different regimes
of complete and partial wetting, and first, second and third layering
transitions located off-bulk coexistence.  Our results for the first
layering binodal line extending into the one-phase gas region are in
qualitative agreement with their findings. Previous comparisons with
DFT results \cite{brader03molphys} left a puzzle because a layering
line was found for $q=0.6$, but not for $q=1$.  The length of this
first layering binodal line obtained from DFT is considerably larger
(in range of $\eta_p^r$) than that found in
simulations~\cite{dijkstra02swet}.  With hindsight, this deviation
seems consistent with that in other situations, as too small values
for the critical polymer fugacities compared to simulations are also
found in bulk (see e.g.~\cite{vink04codef}) or in confinement in
planar capillaries~\cite{schmidt03capc,schmidt04cape}. We have not
obtained results for the higher-$n$ layering binodal lines, so we can
not compare the DFT for $n>1$ efficiently with simulations.
Nevertheless, the overall agreement is good and the DFT seems to
capture all relevant effects: i.e.\ the layering transitions at bulk
coexistence, a layering line extending into the one-phase region and a
wetting transition \cite{brader03molphys}. Given the approximate
nature of the free energy functional and the fact that the surface
phase transitions are governed by tiny free energy differences, this
is quite remarkable.

\begin{figure}[t]
\epsfig{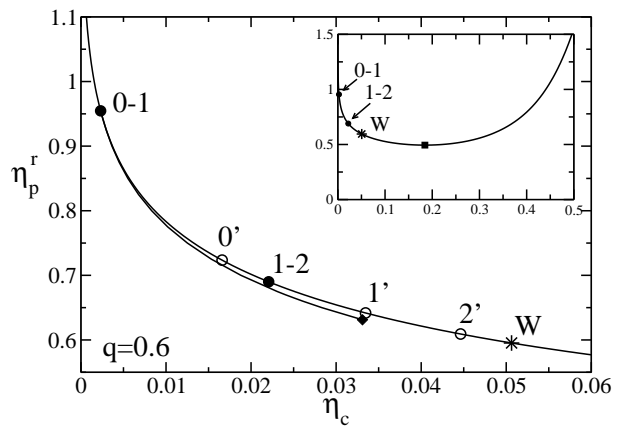}
\caption{\small Same as Fig.~\ref{fig:surfacephaseq1} but for
$q=0.6$. The 0-layer adsorption spinodal
\cite{brader02a,brader03molphys} connecting statepoint $0'$ with the
surface critical point (diamond) is omitted for clarity.}
\label{fig:surfacephaseqp6}
\end{figure}

Coming back to $q=0.6$, we summarize our results and that of Brader et
al.~\cite{brader02a,brader03molphys} in
Fig.~\ref{fig:surfacephaseqp6}. Similar to the case of $q=1$, the
first and second layering transitions at bulk coexistence are located
at higher fugacities than the corresponding 0-layer and 1-layer
adsorption spinodal points, respectively. The effect, however, is more
dramatic than for $q=1$, cf.\ Fig.~\ref{fig:surfacephaseq1}.  Brader
et al.\ have also found a 2-layer adsorption spinodal point,
indicating the existence of a third layering transition, whose binodal
we have not been able to locate.  We have also determined the first
layering line extending into the bulk gas phase and find it to be
remarkably long. The location of the critical point is determined with
the same fit procedure as described above for the first layering
transition for $q=1$.  The large separation along the bulk binodal of
the second layering transition and the 1-layer adsorption spinodal
point suggests again the presence of a second layering line.


\section{Conclusion}

In conclusion, we have investigated the contact angle, $\theta$, of
the (colloidal) liquid-gas interface and a hard wall using the AO
model colloid-polymer mixture and considering two different
polymer-to-colloid size ratios, $q=0.6$ and $q=1$. 
Our results for $\theta$ are obtained via Young's equation
from independent numerical DFT calculations of the
liquid-gas, the wall-gas, and the wall-liquid interfacial free energies
at bulk coexistence.  
At the planar wall-gas interface at bulk coexistence, we identify a range of different
metastable states each corresponding to a number of adsorbed
colloid layers at the wall.
We argue that the globally stable state corresponds to the lowest wall-gas interface tension,
and therefore possesses the largest value of $\theta$.
For small density differences of the coexisting liquid and gas, i.e.\ close to the bulk critical 
point, the wall is completely wet by colloidal liquid~\cite{brader02a,brader03molphys}
and hence, $\theta=0$. 
Moving along bulk coexistence, away from the critical point, 
the wetting transition~\cite{brader02a,brader03molphys} is crossed, and
$\theta$ becomes non-zero. 
However, typical values of $\theta$ remain very small in a considerable range of statepoints in the partial
wetting regime. 
Mediated by a sequence of first-order layering transitions at the wall-gas interface, 
the contact angle grows upon moving further from the
critical point and reaches typical values up to $\sim 35^{\rm o}$.
These layering transitions of the coexisting colloidal gas in contact with the wall
appear as a discontinuities in the slope of $\theta$ as a function of a
thermodynamic control parameter, e.g.\ the polymer reservoir packing
fraction or the colloid packing fraction difference between both
coexisting bulk phases.

Previous difficulties to measure the contact angle accurately~\cite{aarts03} 
have been overcome by Aarts and Lekkerkerker with the
use of confocal scanning laser microscopy~\cite{aarts04codef} in a
system with $q=0.56$.  The authors conclude that $\theta=0$ for all
statepoints considered, consistent with their direct observation of a
prominent colloid wetting film at the interface of the bulk gas with
the wall. They point out that actual values of $\theta$ are very
sensitive to the precise determination of the location of the wall.  Large values of
contact angles as well as the observation of the transition to complete wetting
have been reported by Wijting et al.~\cite{wijting03a,wijting03b},
using extrapolation of dynamical measurements (i.e.\ moving the wall) to zero velocity. 
However, some reservations have been made with respect to the latter results~\cite{aarts04codef}.
The magnitude of the contact angle results from subtle differences between the interface tensions
and we do not expect our present results to resolve experimental issues.
The contact angles which we have calculated for the highly idealized AO model
in contact with a hard wall can therefore only serve as a reference case.
Important effects due to more realistic polymer-polymer interactions~\cite{aarts04cahn},
polydispersity and gravity are not captured in our present model
(see Ref.~\cite{brader03molphys} for a discussion).

We have also reconsidered the surface phase behavior of the AO model
colloid-polymer mixture at a hard wall. 
This system is known to exhibit a sequence of first-order layering transitions 
upon following the gas branch of the liquid-gas bulk binodal towards the bulk critical point
(i.e.\ reducing $\eta_p^r$).
In addition, layering lines extending off-bulk coexistence into the one-phase gas
region have been located.
Such a layering transition is characterized by a jump in the colloid adsorption at the wall 
and can be identified as the growth of an additional colloid layer at the wall-gas
interface~\cite{brader02a,brader03molphys,dijkstra02swet}. 
For one specific case, being the first layering transition for size ratio $q=1$, 
we have determined the layering binodal, which gives the equilibrium location of
the transition, to high accuracy.
In addition, we have located the 0-layer adsorption spinodal line,
beyond which (for higher $\eta_c$ at constant $\eta_p^r$) the 0-layer state is unstable
and the 1-layer desorption spinodal line, marking the end of stability of the 1-layer
state (i.e.\ for lower $\eta_c$ at constant $\eta_p^r$).
The layering binodal and the adsorption and desorption spinodal
lines end at a critical point, located in the single-phase gas region of the bulk
phase diagram.
The crossing point of the layering binodal and the bulk
binodal represents a triple point between the bulk liquid and the two
layered states (0 and 1 layers) of the bulk gas which have different values of the adsorption of both components. 
We find the location of this triple point to
differ substantially from the (previously identified~\cite{brader02a,brader03molphys}) 
crossing point of the adsorption spinodal and
the bulk binodal. 
Remarkably, we could not find a crossing point of
the desorption spinodal and the bulk binodal
and its absence gives rise to continued (meta)stability of the 1-layer state upon
increasing $\eta_p^r$ at coexistence.  
We believe that this scenario holds for higher layering transitions and other size ratios. 
We have presented further results for the second layering transition for $q=1$ as well
as for first and second layering transitions for $q=0.6$.  
The order of the wetting transition and whether it occurs via an infinite or a 
finite sequence of layering transitions remain open questions. 
Whether the occurrence of layering transitions is specific to the AO model (see also
Ref.~\cite{brader03molphys} for a more extensive discussion) or would
be present in more realistic descriptions of colloid-polymer mixtures,
is another interesting question. 
However, any experimental attempt to reveal such (layering) phase behavior would require exceptional accuracy 
for determining $\theta$ or resolution on the particle level for direct observation.

\begin{acknowledgments}
D.\ G.\ A.\ L.\ Aarts is thanked for pointing out the relevance of the
wall contact angle of the colloidal liquid-gas interface to us.  We
acknowledge useful discussions and correspondence with R.\ Evans, J.\
M.\ Brader, and R.\ Roth.  R.\ Evans is also thanked for critically
reading of what turned out to be a preliminary version of our
manuscript and R.\ Roth for an independent check of numerical results.
This work is financially supported by the SFB-TR6 program ``Physics of
colloidal dispersions in external fields'' of the \emph{Deutsche
Forschungsgemeinschaft} (DFG).  The work of MS is part of the research
program of the \emph{Stichting voor Fundamenteel Onderzoek der
Materie} (FOM), that is financially supported by the \emph{Nederlandse
Organisatie voor Wetenschappelijk Onderzoek} (NWO).
\end{acknowledgments}

\bibliographystyle{unsrt}
\bibliography{paulslit}

\end{document}